# A Multimodal Engineering GraphRAG Framework for Post-Wind-Event Building Assessment and Recovery Inference

**Seyedeh Fatemeh Mirfakhar[1], Reda Snaiki[1*], Rocio Lilen Segura[2]**

[1] *Department of Construction Engineering, École de Technologie Supérieure, Université du Québec, Montréal, Québec, Canada*
[2] *Department of Civil, Geological and Mining Engineering, Polytechnique Montréal, 2500 Chemin de Polytechnique, Montréal, QC H3T 0A3, Canada*

[*] *Corresponding author. Email*: reda.snaiki@etsmtl.ca

**Abstract:** Post-event building assessment requires more than identifying physical damage; observed conditions must be related to their consequences for safety, habitability, functionality, and recovery. However, post-disaster evidence is typically heterogeneous, incomplete, and distributed across images, field observations, and reconnaissance reports. Existing recovery models, in contrast, generally assume that damage has already been translated into structured engineering variables. This study develops a multimodal engineering Graph Retrieval-Augmented Generation (GraphRAG) framework for post-wind-event building assessment that connects observed damage to physical consequences, building-condition states, recovery actions, and downtime. Image- and text-based evidence is transformed into a provenance-preserving component-level representation that distinguishes confirmed damage, explicit negative evidence, unresolved conditions, and temporary measures. Graph-aligned retrieval combines comparable building scenarios with explicit engineering relationships, while compatibility, contradiction, evidence-precedence, and abstention mechanisms limit unsupported inference under partial observability. Recovery actions are derived from supported damage–consequence relationships, and downtime uncertainty is propagated through an action-dependency network using Monte Carlo simulation. The framework is demonstrated for tornado-affected low-rise residential buildings using post-event reconnaissance and image-based evidence. Comparison with BaselineRAG and Structured Baseline RAG shows that GraphRAG improves agreement for key building-condition and recovery outcomes while providing traceable engineering reasoning and preserving evidence provenance. Ablation analyses further demonstrate the importance of structured engineering context, evidence safeguards, and graph-path traversal. The results highlight the potential of engineering GraphRAG for transparent, recovery-oriented assessment from incomplete multimodal post-event evidence.

**Keywords:** GraphRAG; post-wind-event assessment; multimodal evidence; building functionality; recovery assessment.

## 1. Introduction

Extreme wind events can produce highly heterogeneous building damage, ranging from localized envelope damage to extensive structural failure, with direct implications for safety, habitability, functionality, and recovery. Post-event assessment therefore requires more than identifying the presence or severity of physical damage: the consequences of an observed condition depend on the affected component, its role within the building system, and whether the building can continue to provide safe occupancy and its intended function. Functional-recovery research has increasingly emphasized this distinction by linking component-level damage to reoccupancy, functionality, and restoration time rather than treating a global damage state as a direct measure of post-event usability [1,2]. Similar component- and functionality-based approaches have been developed for extreme-wind and tornado applications [3-5]. In practice, however, post-event assessments must often be performed from heterogeneous and incomplete evidence, including photographs, field observations, reconnaissance reports, and documented building cases that differ in terminology, spatial coverage, and level of detail [6-8]. A central engineering challenge is therefore to transform this distributed evidence into a consistent representation that connects observed damage to its physical consequences, building condition, functionality, and recovery requirements.

Building functionality and recovery have been modeled through component-based restoration frameworks, empirical recovery observations, and performance-based assessment approaches. These studies relate structural and nonstructural damage to building functionality, repair sequences, and restoration time, allowing partial functionality and progressive recovery to be represented rather than treating recovery as a single transition from damaged to fully restored conditions [2,3,9-11]. Related work has extended these concepts to wind-damaged buildings and community-scale resilience by linking component or building damage to service disruption, shelter capacity, restoration trajectories, and broader recovery consequences [4,5,12,13]. However, these approaches generally begin with damage already expressed through predefined damage states, component-level variables, fragilities, or other structured engineering descriptors. Their application to post-event assessment therefore depends on a preceding step capable of transforming heterogeneous and partially observed field evidence into the structured component-level information required for functionality and recovery inference.

Post-event building conditions are increasingly documented through ground-level photographs, aerial and remote-sensing imagery, and reconnaissance datasets. Recent advances in computer vision and deep learning have enabled automated building localization, segmentation, damage detection, and severity classification from these sources [6,7,14-17]. More recent tornado-focused studies have further connected image-derived damage assessments to restoration and community-recovery models, providing an important bridge between post-event observations and downstream resilience analysis [18,19]. However, these approaches typically convert visual evidence into predefined damage states or ordinal categories, potentially obscuring distinctions such as multiple damaged components with different mechanisms and severities. They may also fail to distinguish confirmed absence from unobserved conditions, temporary protection from underlying physical damage, or conflicting observations across sources. Automated damage recognition therefore does not by itself establish how individual observations relate to physical consequences or how multiple observations collectively constrain building condition. This motivates a representation that preserves component-level engineering context, evidence provenance, and partial observability while explicitly linking observed damage to downstream consequences and recovery.

Retrieval-augmented generation (RAG) provides a natural framework for grounding language-model reasoning in external evidence by retrieving relevant information from a non-parametric knowledge source before response generation [20,21]. Multimodal extensions further enable retrieval across both visual and textual information, with recent applications exploring weather archives, and technical engineering documents for disaster assessment [22-27]. However, semantic similarity does not necessarily imply engineering compatibility [28]. Observations concerning the same building component can be textually similar while representing fundamentally different physical conditions and recovery implications, such as roof-covering loss, roof-structure failure, temporary roof protection, or an explicitly intact roof condition. Likewise, an unobserved component cannot be treated as confirmed undamaged, and contradictory evidence should not be merged as if it were mutually supportive. Conventional RAG alone does not explicitly enforce these distinctions or represent the directed relationships linking damage to physical consequences, building-condition states, functionality, and recovery [28,29]. GraphRAG offers a promising alternative by representing engineering concepts and their relationships explicitly and enabling retrieval through graph structure [27,30]. Yet its application to operational post-event building assessment from incomplete, multimodal, and potentially conflicting evidence remains limited, motivating a framework that combines graph-based engineering relationships with explicit compatibility, contradiction, provenance, and evidence-control mechanisms.

To address these gaps, this study develops a multimodal engineering GraphRAG framework for post-wind-event building assessment and demonstrates its implementation for tornado-affected low-rise residential buildings. The framework transforms image- and text-based evidence into a provenance-preserving component-level representation that distinguishes confirmed damage, explicit negative evidence, unresolved conditions, and temporary measures. It then combines structured scenario retrieval with explicit engineering graph relationships to link observed damage to physical consequences, building-condition states, functionality, recovery actions, and downtime. Compatibility, contradiction, evidence-precedence, and abstention mechanisms are incorporated to limit unsupported inference under partial observability, while semantic similarity is used only as a secondary tie-resolution mechanism. The framework is evaluated using held-out building scenarios, comparisons with BaselineRAG and Structured Baseline RAG, and ablation analyses examining the contributions of engineering context, evidence safeguards, and graph-path traversal. The case study therefore evaluates both predictive agreement with the operational reference states and the added value of traceable, recovery-oriented engineering reasoning.

## 2. Methodology

### 2.1 Framework overview

The framework addresses a common limitation of post-disaster assessment: physical damage is often well documented through images and reconnaissance reports, whereas information on functional consequences and recovery is sparse, fragmented, or reported separately. By integrating multimodal evidence with structured engineering knowledge, the framework supports traceable, recovery-oriented assessment while preserving the provenance and uncertainty of the available evidence. An overview is shown in Fig. 1.

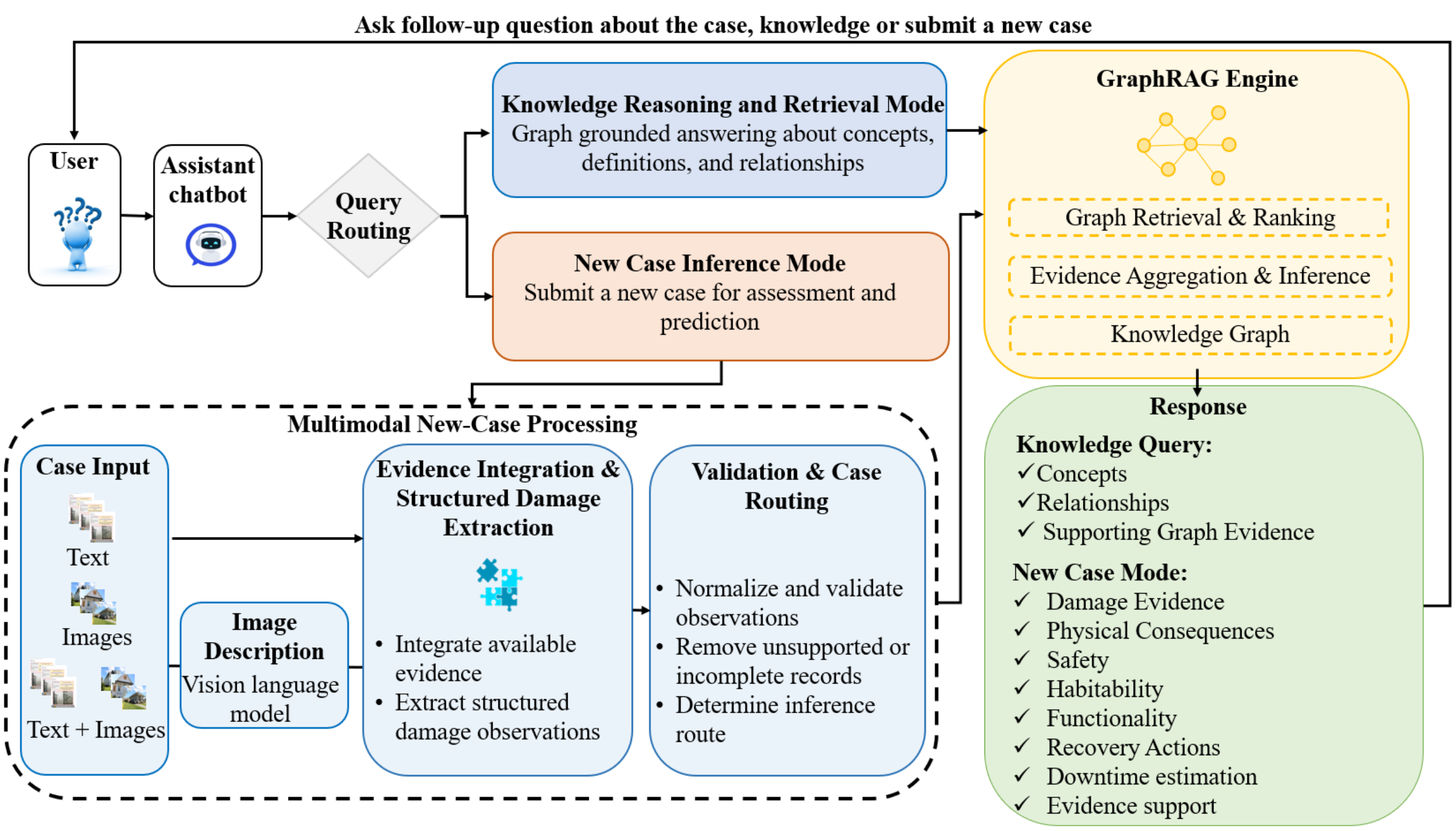

**Fig. 1** Overall architecture and interaction flow of the proposed multimodal GraphRAG framework for post-wind-event building assessment and recovery inference.

The framework operates in two modes: (i) Knowledge reasoning and retrieval mode addresses questions concerning engineering concepts, relationships, and stored scenarios. User queries are routed to the GraphRAG engine, which retrieves relevant graph information and generates responses grounded in the corresponding concepts, relationships, scenarios, and sources. (ii) New case inference mode assesses previously unseen buildings using images, written descriptions, or both. For multi-image cases, the images are treated as complementary evidence after confirmation that they represent the same building. Image inputs are first converted into descriptions of directly observable conditions and integrated with the available textual information. The combined evidence is then transformed into structured observations describing the affected component, damage mechanism, severity, evidence certainty, supporting evidence, and source.

The framework explicitly distinguishes confirmed damage, explicit negative evidence, unresolved conditions, and temporary measures. For example, visible roof-covering loss is treated as confirmed damage, an explicitly observed intact roof covering as negative evidence, and an obscured or indeterminate roof condition as unresolved. A tarp or boarded opening is retained as a temporary measure without assigning the concealed physical condition. Conditions that are not observed or described remain unasserted rather than being treated as undamaged. After validation, the structured case representation is used for graph-aligned retrieval and engineering inference. All inferred outcomes retain links to the evidence and graph relationships supporting them. The current case and inference context are preserved during the conversation with the chatbot assistant, allowing follow-up questions without repeating the complete processing workflow.

### 2.2 Wind-engineering knowledge graph construction

Post-event building damage information may be distributed across heterogeneous sources and levels of assessment. Images and reconnaissance reports can primarily document observable physical damage [31], whereas information on physical consequences, building-condition states, recovery actions, and recovery durations may be available from complementary engineering sources or established through engineering interpretation [2,3]. The proposed knowledge-graph architecture is designed to integrate these different forms of information within a common engineering representation while preserving scenario and source provenance. It provides the structured basis for graph-aligned retrieval and subsequent inference of building condition and recovery.

#### 2.2.1 Knowledge sources and standardized scenario representation

The general framework accommodates two complementary source categories: (i) documented engineering information from reconnaissance reports, technical documents, and research publications, including resources from the Structural Extreme Events Reconnaissance (StEER) Network [32], (ii) visual evidence obtained from post-event building images. Document-based sources may provide building characteristics, observed damage, and available assessment information, whereas image-based sources provide descriptions of directly observable building conditions generated through the vision-language processing stage. Regardless of source type, the available information is organized into a common building-scenario representation. Each scenario $s_{\mathrm{i}}$ is represented as:

$$s_i = \{M_{\mathrm{i}}, B_{\mathrm{i}}, D_{\mathrm{i}}, C_{\mathrm{i}}, S_{\mathrm{i}}, H_{\mathrm{i}}, F_{\mathrm{i}}\,, R_{\mathrm{i}}, A_{\mathrm{i}}, T_{\mathrm{i}}\} \quad (1)$$

where $M_i$ denotes source metadata, $\boldsymbol{B_i}$ building information, $D_{\mathrm{i}}$ the set of component-level damage records, and $C_{\mathrm{i}}$ the associated physical consequences. $S_{\mathrm{i}}$, $H_{\mathrm{i}}$, and $F_{\mathrm{i}}$ denote safety, habitability, and functionality states, respectively, while $R_{\mathrm{i}}$, $A_{\mathrm{i}}$, and $T_{\mathrm{i}}$ represent recovery class, recovery actions, and downtime information.

Each damage record specifies the affected component, damage mechanism, severity, evidence certainty, supporting evidence, and source. For knowledge-base scenarios, Eq. (1) contains the information available from the corresponding source and engineering representation. For a new case, only the available case evidence is initially populated; building-condition and recovery attributes are subsequently inferred by the framework. This standardized representation provides a common interface through which report-based observations and VLM-derived image descriptions can be normalized, validated, and incorporated into the knowledge graph without requiring all attributes to originate from the same evidence pathway.

### 2.2.2 Canonical engineering concepts and controlled vocabulary

A controlled vocabulary standardizes engineering terminology across heterogeneous sources while preserving the original scenario-specific observations. For damage representation, the canonical vocabulary defines component types and families, damage mechanisms and mechanism families, and severity levels. Source terminology is mapped to these concepts through:

$$\phi \;:\; x_{raw} \rightarrow x_{canonical} \quad (2)$$

while the original terminology and its provenance are retained. Canonical concepts enable comparison across scenarios expressed using different terminology. Component and damage-mechanism families group related concepts and define broader admissible matches among them. Belonging to the same family does not make concepts equivalent or automatically interchangeable during retrieval; it only permits a broader match under the defined engineering hierarchy. The representation therefore preserves case-specific evidence while providing the standardized concepts required for cross-scenario retrieval.

### 2.2.3 Knowledge-graph schema, provenance, and engineering relationships

The standardized scenarios are represented as a directed typed knowledge graph:

$$G = (V, E, \tau_{\mathrm{v}}, \tau_{\mathrm{e}}) \tag{3}$$

where $V$ is the set of nodes, $E \subseteq V \times V$, the set of directed relationships, and $\tau_{\mathrm{v}}$ and $\tau_{\mathrm{e}}$ define the corresponding node and relation types. The graph connects three complementary forms of information. The evidence and scenario representation preserves building cases, source metadata, and observed conditions. The engineering-semantic representation connects components and damage mechanisms with severity, physical consequences, and building-condition states. The recovery representation links supported damage and consequences to recovery classes, required actions, and downtime information. These representations are interconnected rather than treated as independent layers, preserving a traceable path from the original evidence to its engineering and recovery implications.

Within each scenario, observed damage is linked to its component and damage mechanism, corresponding canonical concepts, and supported physical consequences. Safety and habitability information is connected to functionality, while damage–consequence relationships provide the basis for recovery actions and recovery information. For example, a scenario-specific observation of partial roof-covering loss retains its original building and source context while being mapped to the canonical roof covering component and loss damage mechanism. It can then be connected to supported consequences such as envelope breach or loss of weather protection and, where applicable, to the corresponding recovery actions.

Scenario identifiers and source metadata are retained throughout these relationships so that retrieved evidence can be traced to its origin. The resulting graph therefore combines case-specific observations, standardized engineering concepts, and recovery information without treating related concepts as equivalent or obscuring the evidence on which they are based. Figure 2 summarizes the knowledge-graph construction and its role in the subsequent retrieval process.

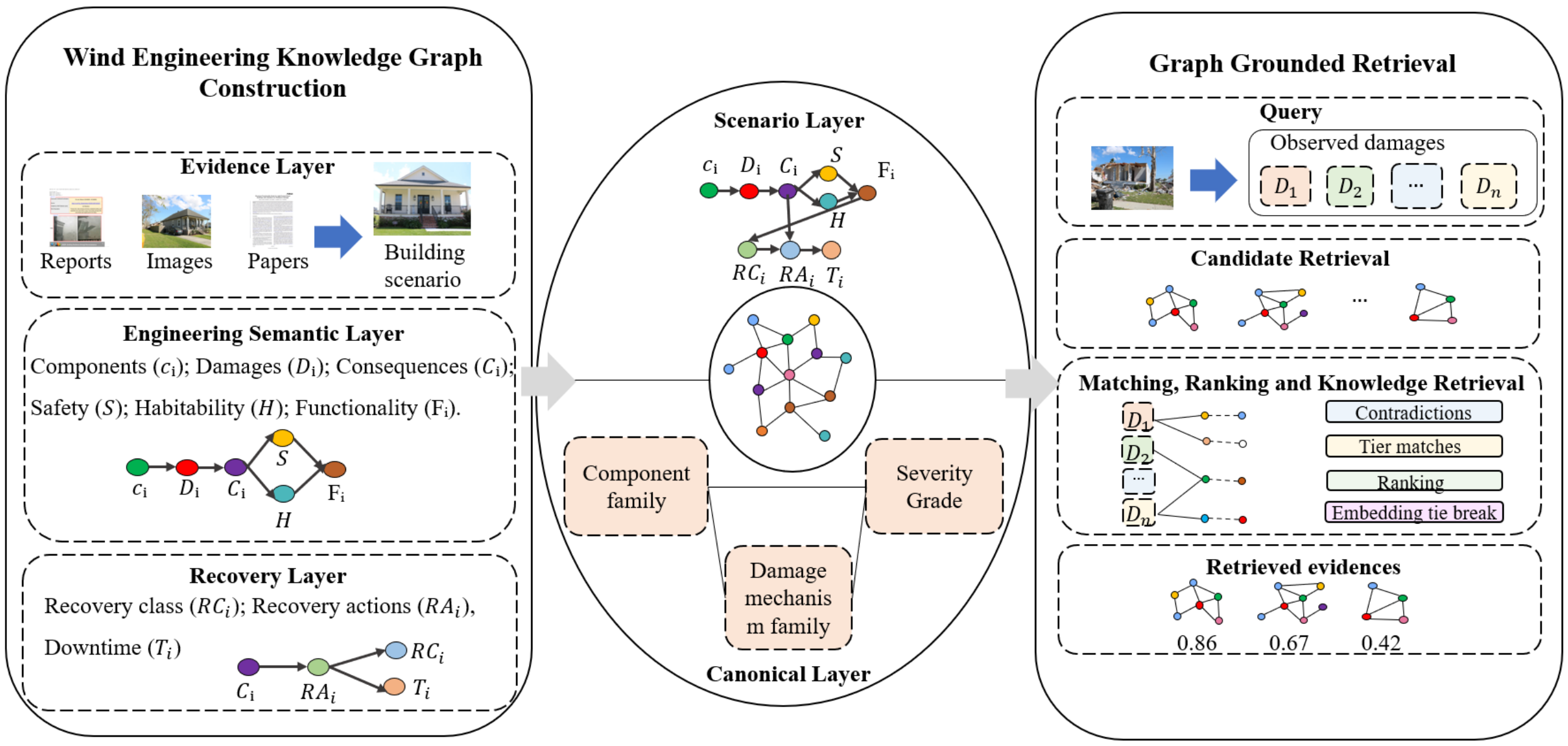

**Fig. 2** Wind-engineering knowledge-graph representation and graph-aligned retrieval process used in the proposed framework.

### 2.3 Multimodal new-case processing

For a new building case, the framework converts the available image and textual evidence into a validated structured query before graph retrieval. At this stage, the framework represents only observable physical conditions and does not assign engineering consequences, building-condition states, or recovery outcomes. Partial observability is explicitly preserved so that unobserved or indeterminate conditions are not interpreted as evidence of absence. The processing sequence comprises multimodal evidence integration, structured observation extraction and validation, and evidence-based query routing.

#### 2.3.1 Multimodal input integration

A new case may contain a written description, one or more building images, or both. When multiple images are provided, they are treated as complementary views only after confirmation that they correspond to the same building. Each image is processed by a VLM to generate a neutral description restricted to directly observable building conditions. Hidden or concealed damage is not inferred from visual cues alone. Image-derived descriptions are then combined with any available textual evidence while preserving the source of each observation. The integrated evidence is subsequently passed to the structured extraction stage.

#### 2.3.2 Structured evidence extraction and validation

The integrated case evidence is converted into a set of structured observations. For a query case $q$, the evidence representation is written as:

$$Q_q = \{D_q^+, D_q^-, U_q, M_q\} \tag{4}$$

where $D_q^+$ denotes confirmed damage observations, $D_q^-$ explicit negative observations, $U_q$ unresolved or unobservable conditions, and $M_q$ observed temporary measures. The four evidence categories are treated differently. A confirmed damage observation provides positive evidence of a physical condition, whereas an explicit negative observation records that a specified condition was observed to be absent. Unresolved conditions correspond to regions or components for which the available evidence is insufficient to establish either the presence or absence of damage. Temporary measures, such as tarps or boarded openings, are recorded separately because they may indicate prior intervention without establishing the concealed physical condition. Each confirmed damage observation records the affected component, damage mechanism, severity, evidence certainty, supporting evidence, and source, with terminology normalized according to the controlled vocabulary defined in Section 2.2.2. Evidence certainty is represented by predefined categories that determine whether an observation can support evidence routing and recovery-action selection; these categories are not converted into numerical weights in the retrieval score. Extracted observations are validated against their supporting evidence and the controlled engineering representation. Unsupported or inconsistent records are removed, and duplicate observations are consolidated. This validation prevents missing observations from being interpreted as no damage and preserves the distinction between directly observed physical conditions and indirect indicators. Validated positive damage observations provide the primary input to damage-based graph retrieval. Negative observations, unresolved conditions, and temporary measures are retained to support contradiction checking, evidence-adequacy assessment, query routing, and downstream inference safeguards.

### 2.3.3 Query routing

Following validation, each case is assigned to an inference route according to the strongest condition supported by the available evidence. The routing procedure distinguishes: global collapse or major structural failure; confirmed component-level damage; temporary-measure-only evidence; explicit no-damage cases; and insufficient-evidence cases. Global-collapse or major-structural-failure routes require explicit evidence of whole-building collapse or severe structural failure. Localized roof, wall, opening, or attached-element damage remains at the corresponding component scope unless broader structural failure is directly supported.

A no-damage route is assigned only when explicit negative evidence provides adequate coverage of the relevant observable building conditions. Failure to extract positive damage is not sufficient to establish a no-damage condition. The evidence-adequacy criterion used for this decision is defined for the corresponding application in the case-study implementation. Cases represented only by temporary measures are routed separately because the underlying physical condition may remain unresolved. Cases for which the available evidence does not support an admissible route result in abstention from downstream inference. The resulting routed case representation forms the structured query used by the graph-aligned retrieval procedure described in Section 2.4.

### 2.4 Graph-aligned evidence retrieval and scenario ranking

Following query routing, the validated case representation is used to retrieve two complementary forms of evidence from the knowledge graph: comparable building scenarios and relevant canonical engineering relationships. These retrieval views remain separate so that scenario-specific evidence is not conflated with general engineering knowledge. Importantly, target outcomes stored for candidate scenarios are excluded from scenario matching and ranking, preventing outcome information from influencing retrieval.

#### 2.4.1 Candidate identification and damage compatibility

Candidate scenarios are identified by comparing each validated positive damage observation in the query with damage records stored in the knowledge graph. Compatibility is evaluated using the canonical component, damage mechanism, and severity defined in Section 2.2.2. Compatibility is classified using the five ordered tiers summarized in Table 1, ranging from the strongest correspondence ($C_1$) to the weakest admissible correspondence ($C_5$). These tiers define an ordinal engineering hierarchy rather than calibrated numerical distances between match types.

**Table 1** Compatibility tiers used for candidate identification and damage matching.

| Compatibility tier | Correspondence |
|---|---|
| $C_1$ | Exact component–mechanism–severity correspondence |
| $C_2$ | Exact component and mechanism with adjacent severity |
| $C_3$ | Exact component with a compatible damage-mechanism family |
| $C_4$ | Related component and damage-mechanism families |
| $C_5$ | Exact component |

For example, for a query observation of major roof-covering loss, a stored observation of major roof-covering loss produces a $C_1$ match, whereas minor roof-covering loss produces a $C_2$ match. An observation such as roof-covering missing may produce a $C_3$ match when missing and loss belong to the same controlled damage-mechanism family. Records with no admissible correspondence are treated as unmatched. A stored scenario is retained as a candidate when at least one of its damage records forms an admissible match with a positive query observation. For multi-damage cases, partial matches are retained at this stage and subsequently evaluated against the complete set of query observations.

### 2.4.2 Damage matching and scenario ranking

Each candidate scenario $s$ is evaluated against the complete set of positive query observations. For each query observation, the strongest admissible match available within $s$ is retained according to the compatibility hierarchy defined above. When multiple candidate records provide the same compatibility tier, provenance is used as a categorical preference: direct observation precedes source-reported evidence, followed by image-derived hypotheses and controlled graph-expanded fallback evidence. Provenance categories are retained explicitly and are not converted into numerical weights. Candidate scenarios are ranked lexicographically according to:

$$\mathbf{r}(s) = (C_s, -N_{1,s}, -N_{2,s}, -N_{3,s}, -N_{4,s}, -N_{5,s}) \tag{5}$$

where $C_s$ is the number of explicit contradictions between the query and scenario $s$, and $N_{k,s}$ is the number of query observations whose strongest match occurs at compatibility tier $C_k$. Smaller values of $\mathbf{r}(s)$, under lexicographic ordering, are preferred. Thus, scenarios with fewer contradictions are ranked first, followed by those with more matches at progressively weaker compatibility tiers. This ordering prevents a candidate from compensating for an explicit contradiction or a weak match through a larger number of lower-priority matches. A contradiction is recorded only when a candidate condition conflicts with confirmed negative query evidence or another explicitly incompatible observed condition. Additional damage present in a stored scenario is not penalized when the corresponding query condition is unresolved, unobserved, or occluded. Unknown evidence therefore contributes neither support nor penalty. The retrieval trace retains the selected compatibility tier, evidence provenance, and records responsible for any identified contradiction.

### 2.4.3 Retrieval selection and complementary graph evidence

The highest-ranked candidates form the scenario-analogue retrieval set. Semantic similarity is used only to resolve candidates with identical structured rankings. For tied scenarios, textual representations of the query and candidate evidence are embedded and compared using cosine similarity:

$$\mathrm{sim}(q, s) = \frac{\mathrm{e}_q^T \mathrm{e}_s}{\|\mathrm{e}_q\| \|\mathrm{e}_s\|} \tag{6}$$

where $e_q$ and $e_s$ denote the corresponding embedding vectors. Semantic similarity is therefore a tie-breaking mechanism rather than a weighted component of the primary ranking procedure. Any remaining tie is resolved deterministically.

In parallel, each positive query observation is used to retrieve canonical graph paths connecting compatible damage conditions to supported physical consequences, building-condition states, and recovery actions. Paths directly supported by the query evidence retain their original provenance, whereas paths obtained through controlled component- or damage-mechanism-family expansion are retained separately as fallback evidence. Fallback paths may supplement the available engineering context but remain distinguishable from directly supported evidence and cannot independently establish a severe building state. Standalone engineering-knowledge records that are not associated with a complete building scenario may also be retrieved according to their encoded engineering scope and source provenance. Scenario analogues, canonical graph paths, and standalone engineering knowledge remain distinct in the retrieval trace and provide complementary evidence to the inference procedure described in Section 2.5. The case-study implementation specifies the number of retrieved scenario analogues and the bounds applied to graph-path retrieval.

### 2.5 Engineering reasoning and inference

Following graph-aligned retrieval, the retrieved scenario analogues, canonical graph paths, and engineering knowledge are used to infer building condition, functionality, and recovery. The inference procedure applies a categorical evidence-precedence policy so that evidence contributes according to its direct applicability, provenance, and source independence. Safety and habitability constrain functionality, while accepted damage–consequence–action relationships determine the applicable recovery requirements and provide the basis for downtime estimation. The inference record preserves the evidence and graph relationships supporting each outcome.

#### 2.5.1 Categorical evidence precedence

For each target outcome, retrieved evidence is evaluated according to its relevance and independence. In descending order of precedence, the framework considers directly supported engineering graph paths, independent engineering knowledge, independent scenario analogues, evidence associated with the same source family as the query, and controlled fallback evidence. Evidence from a lower-precedence category does not override admissible evidence from a higher category. To avoid treating repeated information as independent confirmation, only one representative from a given source family is retained at the active evidence level. Consistent evidence contributes to inference, whereas unresolved conflicts reduce the admissible support available for the corresponding outcome and may result in an insufficient-evidence decision. For traceability, the active evidence category, contributing source families, and provenance are retained in the inference record.

### 2.5.2 Building-condition and functionality inference

Safety and habitability are inferred from the admissible evidence and the engineering relationships represented in the knowledge graph. Functionality is then assigned from the resulting safety–habitability combination according to the operational relationships defined for the application, thereby maintaining consistency among safety, habitability, and functionality. The inference preserves the spatial and structural scope of the observed damage. Whole-building collapse or equivalent severe states require direct evidence of global collapse or severe structural failure. Localized roof, wall, opening, or other component damage remains at the corresponding scope unless additional evidence supports broader structural impairment. Severe building states therefore cannot be established solely from loosely analogous scenarios or fallback evidence. Evidence adequacy is also required before assigning states representing the absence of impairment. Lack of observed damage does not by itself establish an unimpaired building condition, and unobserved or concealed regions remain unresolved. Similarly, temporary protection may support applicable recovery or precautionary actions but does not establish the concealed physical condition. When the admissible evidence is insufficient to support a state, the framework abstains rather than assigning an unsupported outcome.

### 2.5.3 Recovery-action and downtime inference

Recovery actions are derived from application-specific damage–consequence–action relationships and the validated evidence for the query case. The specific mappings, action dependencies, and duration parameters are defined for each application through its corresponding engineering rule set. Actions directly supported by confirmed damage remain distinguishable from precautionary actions and actions associated with fallback hypotheses. The framework therefore infers recovery requirements from the observed case rather than transferring the complete recovery information of an analogous scenario. The recovery class is assigned from the supported core recovery actions according to the operational definitions established for the application. Supporting activities may form part of the recovery sequence without independently determining the recovery class. Functionality serves as a consistency constraint on the resulting recovery classification rather than as its direct determinant. Downtime is estimated from the duration distributions and dependency relationships associated with the selected recovery actions.

An engineering dependency network represents action precedence, while actions without dependency constraints may proceed in parallel. For Monte Carlo realization $b$, the start and completion times of recovery action $a$ are defined as:

$$S_a^{(b)} = \begin{cases} 0, & \mathrm{Pred}(a) = \emptyset \\ \max\limits_{p \in \mathrm{Pred}(a)} C_p^{(b)}, & \text{otherwise}, \end{cases} \qquad C_a^{(b)} = S_a^{(b)} + D_a^{(b)} \tag{7}$$

where $\mathrm{Pred}(a)$ denotes the set of predecessor actions, and $S_a^{(b)}$, $D_a^{(b)}$ and $C_a^{(b)}$ are the start time, sampled duration, and completion time of action $a$, respectively. This formulation allows independent actions to proceed in parallel while enforcing the required precedence relationships. The total downtime for realization $b$ is determined by the completion time of the governing terminal action:

$$T^{(b)} = \max_{a \in \mathcal{A}_{term}} C_a^{(b)} \tag{8}$$

where $\mathcal{A}_{term}$ is the set of terminal actions in the recovery network. Repeated simulation produces a downtime distribution from which representative estimates and uncertainty intervals are obtained. The recovery classes, action library, dependency relationships, duration parameters, simulation settings, and reported downtime intervals are specified in the case-study implementation.

### 2.5.4 Traceability and response generation

Traceability is maintained by retaining the validated case evidence, retrieved scenario analogues, canonical graph paths, evidence provenance, and engineering constraints applied during inference. For each predicted outcome, the inference record identifies the evidence category used, the contributing source information, and the graph relationships supporting the decision. Recovery outputs additionally retain the selected actions, their dependency relationships, and the information used for downtime estimation. Directly supported evidence remains distinguishable from fallback evidence throughout the reasoning process. Conclusions relying on fallback support are identified accordingly, and insufficient admissible evidence results in abstention rather than an unsupported state assignment. The conversational response is generated from this inference record and presents the requested outcome together with a concise explanation of its supporting evidence and graph relationships. Follow-up questions concerning the same building use the retained case and inference context without repeating the complete evidence-processing and inference workflow. The case-study implementation and evaluation of the proposed framework are presented in Section 3.

## 3. Case Study

The proposed framework is applied to post-tornado assessment of low-rise residential buildings. This case study was selected because post-event sources provide substantial building-level evidence of physical damage, whereas information on functionality, recovery actions, and downtime is less consistently documented. The case study demonstrates the general framework using tornado-specific building scenarios, operational engineering definitions, and recovery rules. It also evaluates the framework's capabilities in terms of knowledge retrieval and new case inference.

### 3.1 Tornado case-study scope and knowledge-base instantiation

#### 3.1.1 Database and building scenarios

For the tornado case study, the scenario database was assembled from post-event reconnaissance reports and associated datasets produced by the StEER [33–37], complemented by relevant research publications [38–41]. These sources provide building-level photographs, damage observations, construction information, and, where available, reported assessment information. Image-based scenarios were obtained from StEER datasets, whereas document-derived scenarios were extracted from report sections and research publications using building descriptions, figure captions, and associated assessment information. As a proof-of-concept application, the study focuses on tornado-affected low-rise residential buildings, defined here as buildings up to three stories above grade. Cases were retained when they represented an identifiable low-rise residential building with sufficient evidence to characterize its exterior condition. Multiple images or document fragments referring to the same physical building were assigned to a common scenario group, and all records within that group were retained within the same data split (retrieval, validation, or final test) to prevent information leakage across splits. The resulting database contains 581 unique building scenarios from six tornado events, comprising 542 image-based and 39 document-derived scenarios. The database spans conditions ranging from no confirmed visible exterior damage to localized and multi-component damage and whole-building collapse. Although this scope limits variation in hazard type, occupancy, and building context, it was intentionally adopted to provide a controlled proof-of-concept setting while retaining substantial variation in observed damage severity and extent.

#### 3.1.2 Operational definitions and engineering rules

Post-event reconnaissance sources primarily document observed damage and building performance rather than necessarily providing a complete and consistently defined assessment of safety, habitability, functionality, recovery requirements, and downtime for every building. The case-study knowledge base therefore incorporates operational definitions and engineering rules to interpret the available evidence consistently. Safety and habitability definitions are informed by established post-disaster building-evaluation guidance, including the Federal Emergency Management Agency (FEMA) P-2055 guidance and the Applied Technology Council (ATC) ATC-45 field manual [42,43], adapted to the low-rise residential and tornado scope considered in this study. Functionality and recovery definitions draw on building-performance and restoration frameworks [2,3], whereas the specific state mappings and provisional duration assumptions are defined for the present case study.

Building condition is characterized through safety and habitability, while functionality represents the resulting level of building usability and recovery characterizes the required level of intervention. The operational states adopted for the tornado case study are summarized in Table 2.

**Table 2** Operational states used in the tornado case study.

| **Dimension** | State 0 | State 1 | State 2 | State 3 | State 4 |
|---|---|---|---|---|---|
| Safety | $S_0$: Safe | $S_1$: Restricted use | $S_2$: Unsafe | - | - |
| Habitability | $H_0$: Habitable | $H_1$: Limited habitability | $H_2$: Not habitable | - | - |
| Functionality | $F_0$: Fully functional | $F_1$: Limited functionality | $F_2$: Partially functional | $F_3$: Severely impaired | $F_4$: Nonfunctional |
| Recovery | $R_0$: No action | $R_1$: Minor repair | $R_2$: Moderate repair | $R_3$: Major repair | $R_4$: Reconstruction |

Functionality is assigned from the inferred safety–habitability combination according to predefined operational relationships. Specifically, $S_0$–$H_0$ corresponds to $F_0$ (fully functional), $S_0$–$H_1$ to $F_1$ (limited functionality), $S_1$ combined with either $H_0$ or $H_1$ to $F_2$ (partially functional), $S_0$ or $S_1$ combined with $H_2$ to $F_3$ (severely impaired), and $S_2$–$H_2$ to $F_4$ (nonfunctional). Recovery is treated separately from this mapping. During new-case inference, the recovery class is assigned from the core recovery actions supported by the observed damage and associated physical consequences, while functionality serves as a consistency constraint rather than a direct determinant.

The recovery-action library includes component-specific inspection, stabilization, temporary protection, repair, replacement, envelope restoration, and reconstruction. Applicable actions are assigned from the case study damage–consequence–action relationships by considering the affected component, damage mechanism and severity, and the resulting physical consequence. The selection and interpretation of intervention types were informed by relevant wind-vulnerability and post-disaster repair guidance, including guidance on temporary protection and the repair or replacement of damaged roof and envelope components [42,44,45]. For example, roof-covering loss may require restoration of weather protection followed by roof-covering repair or replacement, whereas roof-structure failure may require structural stabilization and structural repair or replacement. Core actions determine the recovery class, whereas supporting or precautionary actions may form part of the recovery sequence without independently changing that class. Action dependencies are retained so that prerequisite actions precede dependent repairs, while actions without a defined dependency may proceed in parallel.

Each recovery action is assigned a lognormal duration distribution characterized by a median duration and logarithmic standard deviation. The resulting action-dependency network is evaluated using the recovery-network formulation defined in Section 2.5.3. For each case, 12,000 Monte Carlo realizations are used to propagate action-duration uncertainty through the recovery network, and the resulting downtime distribution is summarized by its P10, P50, and P90 values. The number of realizations was selected through a convergence analysis across nine representative recovery networks using 20 random seeds per network. A sample size of 12,000 was the smallest tested value for which P10, P50, and P90 were within the larger of 0.1 week or 2% of an independent 600,000-realization reference in at least 19 of 20 runs for every network. Larger sample sizes further reduced sampling variability but did not materially change the reported percentiles.

### 3.1.3 Knowledge-graph construction

The standardized case-study records were implemented as a directed knowledge graph. Across the 581 building scenarios, the resulting graph contains 15,196 nodes and 45,183 directed relationships, comprising a canonical engineering representation of 562 concepts and 809 relationships and 14,634 scenario-specific nodes with 44,374 relationships.

Scenario-specific nodes represent building cases, damage observations, physical consequences, building-condition states, functionality, and recovery information, while canonical concepts provide the standardized engineering representation required for cross-scenario retrieval. For the low-rise residential application, building evidence is organized around the principal exterior systems, including roof systems, exterior walls, openings, attached elements, and overall structural condition. Where supported by the evidence, these systems are resolved into more specific components such as roof covering, roof deck, roof structure, windows, and doors.

Localized component damage is represented separately from building-level conditions such as partial or complete collapse, thereby preserving the spatial and structural scope of the observation. Scenario-specific records are linked to their corresponding canonical component and damage concepts and retain their source provenance. The instantiated graph therefore applies the general schema of Section 2.2 to tornado-damaged residential buildings while maintaining a common engineering representation across all scenarios.

### 3.2 Framework implementation and evaluation protocol

The case-study framework was implemented in Python using NetworkX for knowledge-graph construction and traversal. Building images were processed using Qwen2.5-VL-3B-Instruct to generate descriptions restricted to visible exterior conditions, using deterministic decoding, a maximum generation length of 350 tokens, and half-precision inference. Written descriptions and image-derived descriptions were subsequently processed using Llama 3.2:3B through Ollama to extract the structured observations defined in Section 2.3. Graph retrieval used the component, damage-mechanism, severity, contradiction, and provenance information described in Section 2.4, while all-MiniLM-L6-v2 embeddings were used only to resolve ties between scenarios with identical structured retrieval ranks. The retrieval configuration retained the top 10 scenario analogues and canonical graph paths bounded by a maximum depth of three directed hops from each matched canonical root node.

The 581 scenarios were divided into 70% retrieval, 15% validation, and 15% final-test partitions using grouped partitioning at the physical-building or scenario-group level. Multiple images, descriptions, or document fragments associated with the same building were therefore retained within the same partition. The retrieval partition formed the knowledge base available during inference, the validation partition was used for development-stage configuration and robustness assessment, and the final-test partition remained isolated from implementation decisions. Reference safety, habitability, functionality, recovery, and downtime information for held-out cases was excluded from retrieval and accessed only after inference.

Performance was assessed at the building-condition and recovery levels. Safety, habitability, and functionality were evaluated using accuracy, macro-F1, prediction coverage, and selective accuracy. Macro-F1 was computed as:

$$F1_{\text{macro}} = \frac{1}{C}\sum_{c=1}^{C}\frac{2P_cR_c}{P_c+R_c} \tag{9}$$

where $C$ is the number of classes, and $P_c$ and $R_c$ are the precision and recall for class $c$, respectively. Prediction coverage was defined as the proportion of cases for which the framework returned a safety, habitability, or functionality state rather than abstaining, while selective accuracy was computed over these non-abstained cases. Recovery classification and downtime were assessed against the operational engineering rules defined in Section 3.1.2; these results therefore quantify agreement with the case-study reference framework rather than validation against independently observed post-event recovery outcomes. Downtime performance was additionally characterized using week-based error measures, rank correlation, and coverage of the predicted P10–P90 interval.

GraphRAG was compared with two non-graph retrieval-augmented baselines using the same scenario, held-out queries, structured-damage extraction procedure, number of retrieved candidates, downstream inference rules, and recovery model. BaselineRAG retrieved scenarios directly from unstructured evidence text using all-MiniLM-L6-v2 embeddings and cosine similarity, whereas Structured BaselineRAG applied the same vector-based retrieval to the standardized evidence representation used by GraphRAG. The comparison between Structured BaselineRAG and GraphRAG therefore isolates the added contribution of graph-aligned retrieval and explicit engineering relationships. Event dependence and framework sensitivity are examined separately through the leave-one-event-out ablation analysis presented in Section 3.5.

### 3.3 Representative framework demonstrations

The two operating modes of the proposed framework are demonstrated using representative tornado-residential cases. Knowledge Reasoning and Retrieval Mode illustrates graph-grounded access to stored engineering knowledge and scenarios, whereas New-Case Inference Mode demonstrates the complete assessment of previously unseen buildings from multimodal evidence.

#### 3.3.1 Knowledge Reasoning and Retrieval Mode

Knowledge Reasoning and Retrieval Mode supports queries concerning represented building components and damage mechanisms, physical consequences, building-condition definitions, recovery information, stored scenarios, and evidence provenance.

Figure 3 presents two representative queries illustrating complementary uses of the knowledge graph. The first query, 'What consequences can roof-covering loss produce?', retrieves canonical engineering relationships between the observed damage condition and its supported physical consequences. The second query, 'Which scenarios contain both roof and exterior-wall damage? Provide two examples', retrieves building scenarios containing the specified multi-component damage configuration and traces the examples to their supporting evidence.

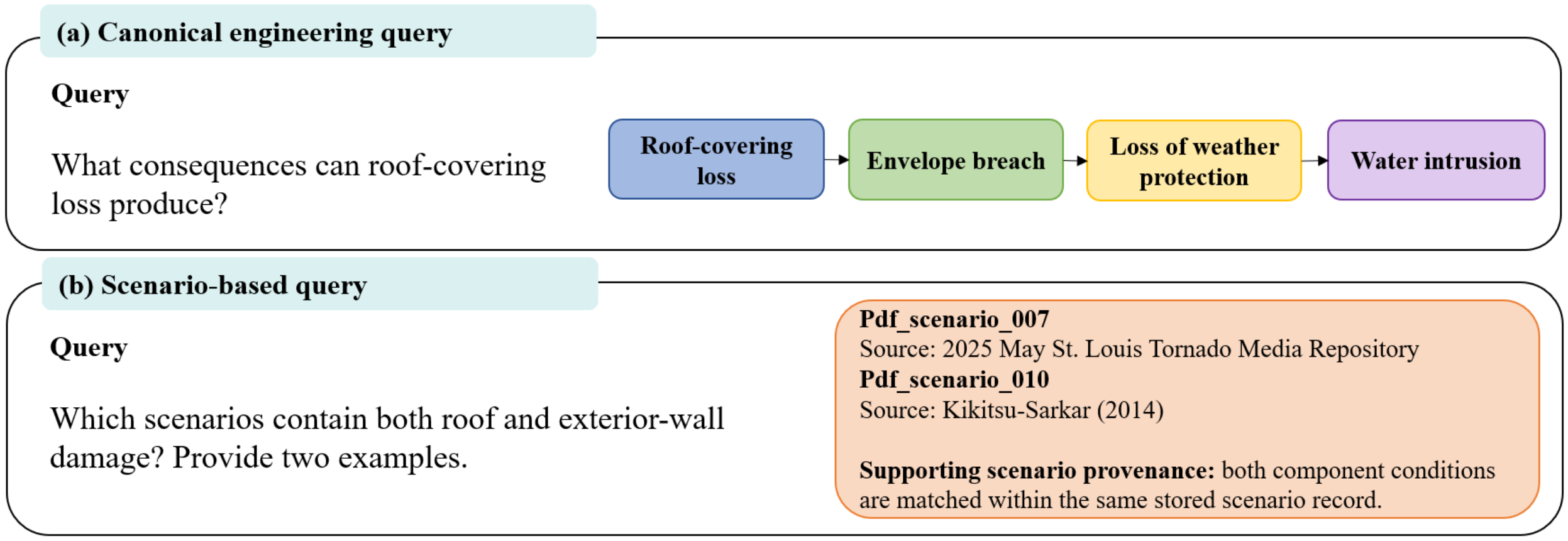

**Fig. 3** Representative graph-grounded knowledge queries illustrating (a) retrieval of canonical damage–consequence relationships and (b) retrieval of stored scenarios with a specified multi-component damage configuration.

### 3.3.2 New-Case Inference Mode

New-Case Inference Mode applies the complete assessment workflow to previously unseen buildings using image evidence, written descriptions, or both. The available evidence is converted into validated structured observations and routed according to the evidence conditions defined in Section 2.3. For cases requiring retrieval, compatible scenario analogues and engineering graph paths are then used to support building-condition, functionality, and recovery inference.

Figure 4 illustrates the two boundary conditions of the routing hierarchy. Figure 4(a) represents a building for which the available exterior evidence provides adequate coverage and no confirmed visible exterior damage. Within the defined exterior-assessment scope, the framework assigns the corresponding unimpaired functionality state and no-action recovery class. This classification reflects only the available evidence and does not constitute a substitute for an on-site structural or occupancy assessment. Figure 4(b) represents the opposite boundary condition, where explicit evidence of whole-building collapse activates the global-collapse route, resulting in a nonfunctional state and reconstruction-level recovery requirement.

**(a) No confirmed visible exterior damage** **(b) Whole-building collapse**

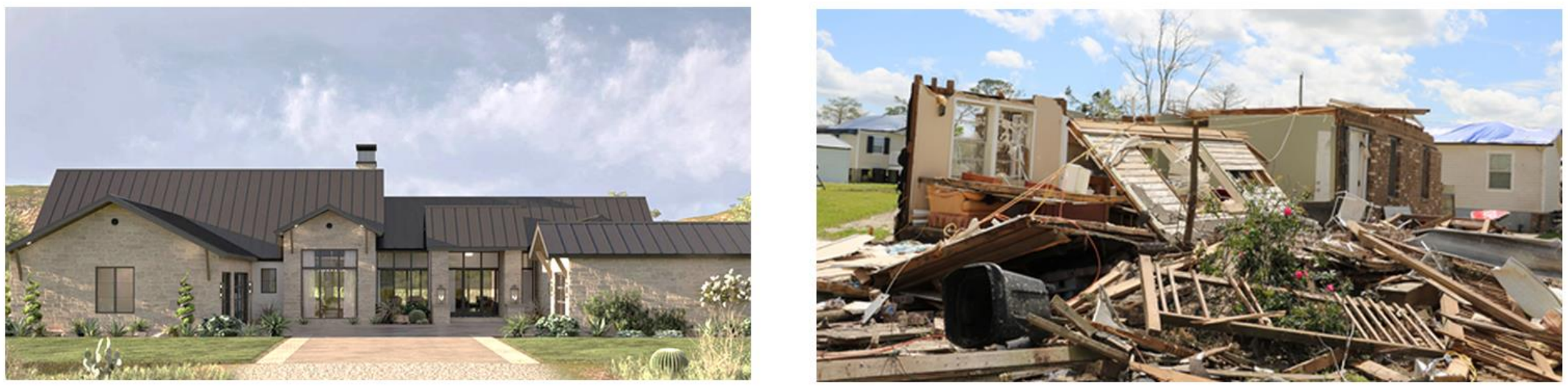

| Assessment output | (a) No confirmed visible exterior damage | (b) Whole-building collapse |
|---|---|---|
| Evidence condition | Adequate exterior evidence; no confirmed visible damage | Whole building collapse |
| Functionality | $F_0$ | $F_4$ |
| Recovery class | $R_0$ | $R_4$ |
| Estimated downtime | 0.0 weeks | P10: 51.3 weeks - Median 82.1 weeks -P90:143.7 weeks |

**Fig. 4** Boundary-condition demonstrations of New-Case Inference Mode: (a) adequate exterior evidence with no confirmed visible damage and (b) explicit whole-building collapse.

Figure 5 presents a more demanding multi-component case requiring graph-supported inference rather than direct boundary routing. The validated observations identify damage involving the exterior wall and roof structure together with an observed boarded opening. Importantly, the boarded opening is retained as a temporary measure, while the concealed opening condition remains unresolved unless independently supported by visible evidence. The directly supported damage observations are linked through the graph to applicable physical consequences, including envelope breach, localized roof instability, and loss of weather protection, which contribute to the resulting safety, habitability, and functionality assessment. Recovery actions are subsequently derived from the accepted damage–consequence–action relationships rather than transferred directly from a retrieved analogue. Their dependency structure is propagated through the Monte Carlo recovery model to obtain the downtime distribution shown in Fig. 5. The reported P10, P50, and P90 values therefore represent uncertainty arising from the recovery-action durations and dependencies defined for the case study.

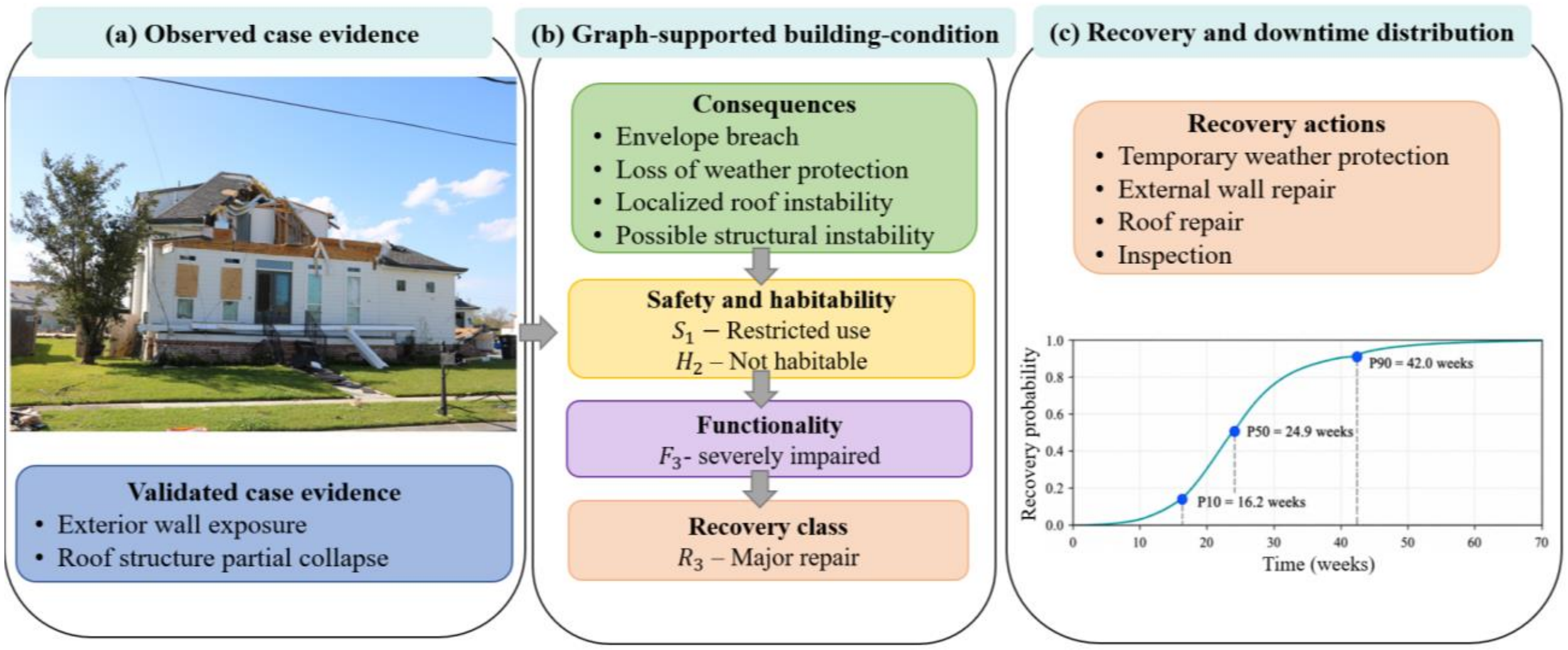

**Fig. 5** Representative multi-component New-Case Inference Mode assessment showing (a) Observed case evidence, (b) graph-supported building-condition and recovery inference, and (c) recovery actions and resulting downtime distribution.

### 3.4 BaselineRAG comparison

The aggregate baseline comparison follows the common evaluation protocol described in Section 3.2. Because Structured BaselineRAG and GraphRAG use the same standardized evidence representation, their comparison more directly isolates the contribution of graph-aligned retrieval and explicit engineering relationships. BaselineRAG provides an additional reference based on retrieval from unstructured evidence text.

Figure 6 presents a comparison between Structured BaselineRAG and GraphRAG using identical multi-view building evidence. The exterior views provide complementary observations, including a boarded opening, temporary roof protection, and additional roof-level evidence, which are integrated into a common structured case representation before retrieval. The corresponding assessment outcomes are summarized in Table 3. Structured BaselineRAG commits to the analogue-based assessment $(S_0; H_1; F_1)$ with inspection required, whereas GraphRAG preserves alternative evidence-dependent outcomes because the condition concealed by the temporary roof protection remains unresolved. If inspection confirms roof-covering damage, GraphRAG retains $(S_0; H_1; F_1)$ but assigns moderate repair. If major localized roof-structure damage is confirmed, the assessment changes to $(S_1; H_2; F_3)$ with major repair and a substantially longer downtime estimate. Structured BaselineRAG relies on states and recovery information associated with semantically similar scenarios, whereas GraphRAG explicitly connects the validated observations to supported engineering relationships. In particular, temporary roof protection is retained as a temporary measure rather than confirmation of the concealed roof condition. GraphRAG consequently distinguishes directly supported recovery actions from precautionary or conditional actions that depend on confirmation of unresolved conditions. The example therefore illustrates the added traceability and evidence control provided by graph-aligned reasoning, as well as its ability to preserve conditional assessment branches when the available observations do not establish the underlying damage condition.

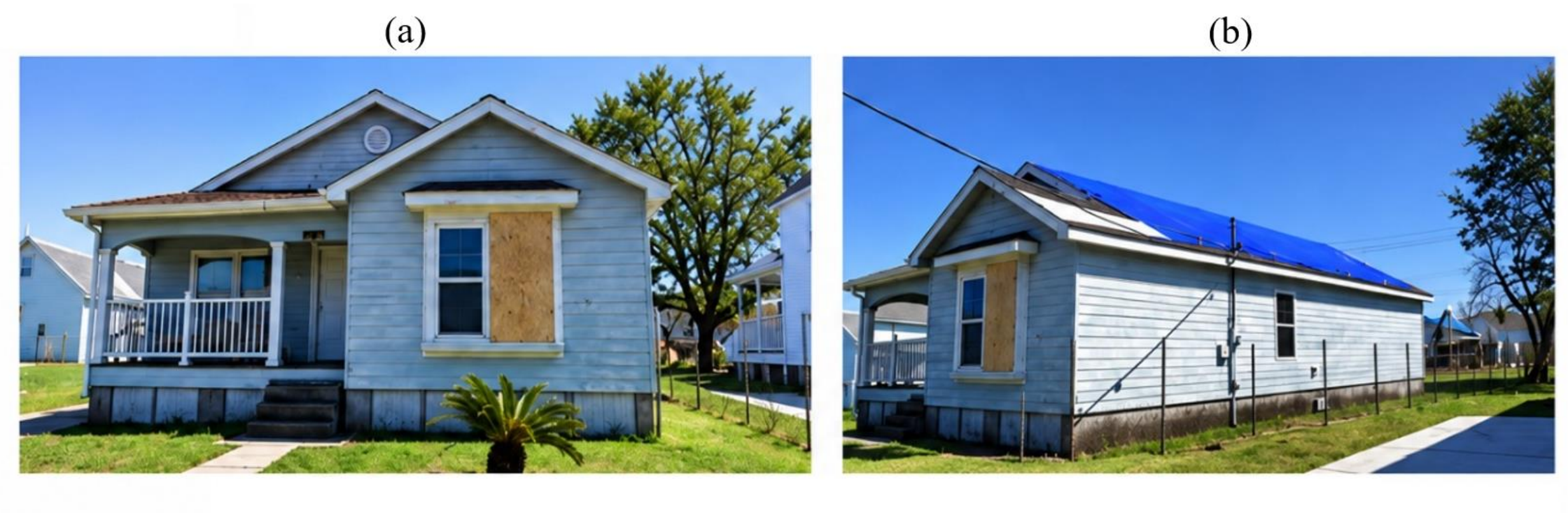

**Fig. 6** Multi-view building evidence used for the representative Structured BaselineRAG–GraphRAG comparison.

The two exterior views provide complementary evidence, including a boarded opening, temporary roof protection, and evidence concerning the roof condition. These observations are integrated into a common structured representation before retrieval.

**Table 3** Case-level comparison of Structured BaselineRAG and GraphRAG for the multi-view evidence shown in Fig. 6.

| Method / condition | Building-condition states | Recovery | Downtime |
|---|---|---|---|
| Structured BaselineRAG | $S_0$, $H_1$, $F_1$ | Inspection required | P10: 3.1; P50: 4.7; P90: 7.4 weeks |
| GraphRAG: roof-covering damage confirmed | $S_0$, $H_1$, $F_1$ | $R_2$: Moderate repair | P10: 3.9; P50: 5.5; P90: 7.8 weeks |
| GraphRAG: major localized roof-structure damage confirmed | $S_1$, $H_2$, $F_3$ | $R_3$: Major repair | P10: 12.6; P50: 20.2; P90: 34.7 weeks |

Structured BaselineRAG commits to the analogue-based assessment, whereas GraphRAG preserves alternative evidence-dependent outcomes because the condition concealed by the temporary roof protection remains unresolved. If inspection confirms roof-covering damage, GraphRAG retains $S_0$, $H_1$, and $F_1$ but assigns moderate repair. If major localized roof-structure damage is confirmed, the assessment changes to $S_1$, $H_2$, and $F_3$, with major repair and a substantially longer downtime estimate.

Structured BaselineRAG relies on states and recovery information associated with semantically similar scenarios, whereas GraphRAG explicitly connects the validated observations to supported engineering relationships. In particular, temporary roof protection is retained as a temporary measure rather than confirmation of the concealed roof condition. GraphRAG consequently distinguishes directly supported recovery actions from precautionary or conditional actions that depend on confirmation of unresolved conditions. The example therefore illustrates the added traceability and evidence control provided by graph-aligned reasoning, as well as its ability to preserve conditional assessment branches when the available observations do not establish the underlying damage condition.

Aggregate performance was evaluated on the held-out scenarios using the same retrieval knowledge base, input evidence, number of retrieved candidates, downstream inference procedure, recovery model, and operational reference definitions across all three approaches. Reference outcomes were excluded from retrieval and introduced only after inference for evaluation. The reported classification metrics therefore quantify agreement with the operational reference states defined for this case study rather than with independently adjudicated post-event building outcomes.

Figure 7 compares macro-F1 and accuracy for safety, habitability, functionality, recovery classification, and categorical downtime. GraphRAG achieves the highest Macro-F1 and accuracy for safety, functionality, recovery, and downtime, while its performance for habitability is approximately the same as Structured BaselineRAG. The largest improvements over Structured BaselineRAG are observed for recovery and downtime, for which macro-F1 increases from 0.918 to 1.00 and from 0.721 to 1.00, respectively. The comparison with Structured BaselineRAG is particularly informative because both approaches use the same standardized evidence representation; differences between them therefore reflect the additional contribution of graph-aligned retrieval and explicit engineering relationships. The results indicate that the principal benefit of GraphRAG occurs for outcomes that depend more directly on connecting observed damage to recovery relationships and downstream consequences. For outcomes with comparable performance, GraphRAG retains the additional capability to trace predictions to the corresponding evidence, engineering relationships, and provenance.

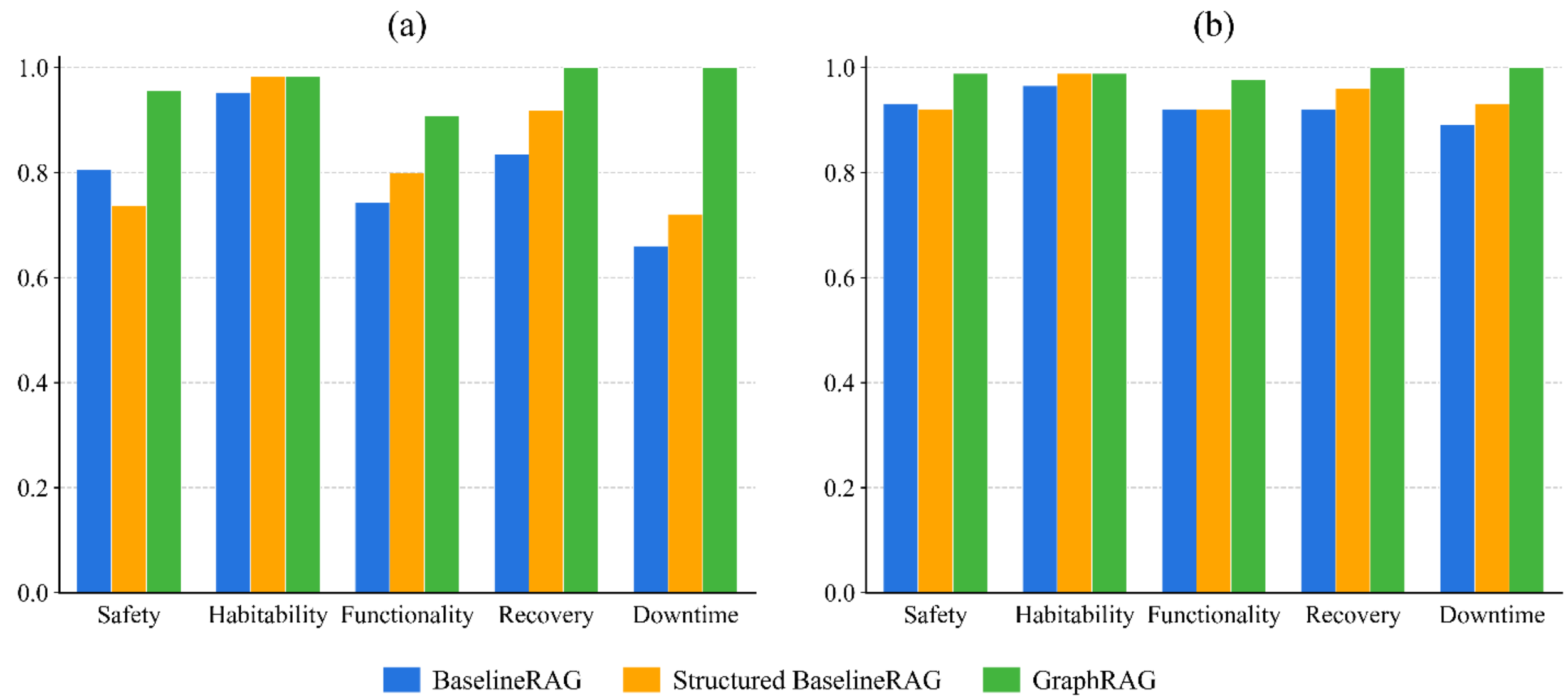

**Fig. 7** Comparison of BaselineRAG, Structured BaselineRAG, and GraphRAG performance across the held-out scenarios: (a) macro-F1 and (b) accuracy for safety, habitability, functionality, recovery classification, and categorical downtime.

Prediction coverage was 100% for all three approaches across the evaluated safety, habitability, and functionality outcomes. Consequently, no abstentions occurred in these evaluation cases, and selective accuracy was identical to the corresponding accuracy values reported in Fig. 7(b).

Continuous downtime performance provides a complementary assessment because recovery duration is estimated on a continuous time scale before being assigned to categorical downtime classes. Table 4 reports the mean absolute error (MAE), median absolute error, root-mean-square error (RMSE), Spearman rank correlation, and P10–P90 interval coverage for the three approaches.

**Table 4** Continuous downtime-estimation performance across the held-out scenarios.

| Method | MAE (weeks) | Median AE (weeks) | RMSE (weeks) | Spearman correlation | P10-P90 coverage |
|---|---|---|---|---|---|
| BaselineRAG | 2.97 | 0.0 | 8.45 | 0.977 | 76.4% |
| Structured BaselineRAG | 1.44 | 0.0 | 5.35 | 0.992 | 93.1% |
| GraphRAG | 0.02 | 0.0 | 0.20 | 1.000 | 99.4% |

The continuous results complement the categorical downtime assessment. Lower MAE and RMSE indicate closer agreement with the reference recovery durations, while Spearman correlation characterizes consistency in the ranking of cases and P10–P90 coverage assesses agreement with the uncertainty intervals generated by the recovery model. GraphRAG achieved the lowest MAE and RMSE and the highest P10–P90 coverage, indicating the closest overall agreement with the model-derived reference recovery durations. GraphRAG also produced the highest Spearman correlation, followed by Structured BaselineRAG, while both structured approaches reduced estimation errors relative to BaselineRAG. Differences between categorical and continuous performance can occur when recovery-duration estimates lie near the boundaries of the downtime classes. Importantly, these metrics quantify agreement with the case-study recovery model and its prescribed action-duration relationships rather than empirical validation against independently observed building-level recovery times.

### 3.5 Ablation and robustness analysis

An ablation analysis was conducted to examine the contribution of three principal components of the GraphRAG framework: graph-path traversal, engineering context, and evidence safeguards. Graph-path traversal represents the explicit use of engineering paths linking observed damage to physical consequences, building-condition states, and recovery information. Engineering context represents the controlled information describing component type, damage mechanism, severity, extent, structural scope, consequences, and recovery relationships. Evidence safeguards represent the controls associated with explicit negative evidence, temporary measures, unresolved conditions, compatibility, and contradiction handling.

Each component was removed independently while the remaining framework configuration, scenario data, operational definitions, recovery model, and evaluation procedure were held fixed. The analysis used six leave-one-event-out outer folds across all 581 scenarios. In each fold, scenarios from one tornado event were used for evaluation, while retrieval and model training used scenarios from the remaining five events. Hyperparameters were selected through an inner event-level validation procedure in which each of the five outer-training events was successively reserved for validation, and the remaining four were used for training. The selected settings were then reused for the complete and ablated configurations without additional tuning, and the reported metrics were calculated from the pooled predictions across the six outer folds.

Figure 8 compares the accuracy and macro-F1 of the complete framework with the three ablated configurations. The relative effect of removing a component depends on both the outcome and the evaluation metric. Removing engineering context produces the largest macro-F1 reductions for safety and functionality, with functionality macro-F1 decreasing from 0.824 to 0.434. Removing evidence safeguards reduces functionality macro-F1 to 0.586 and produces the largest habitability macro-F1 reduction, from 0.984 to 0.602. Removing graph-path traversal also reduces performance, with functionality macro-F1 decreasing to 0.751. The accuracy results show a different pattern. Removing evidence safeguards produces the largest reductions for safety, functionality, and categorical downtime, with accuracy decreasing from 0.902 to 0.579, from 0.893 to 0.571, and from 0.997 to 0.413, respectively. Removing engineering context produces the largest habitability accuracy reduction, from 0.983 to 0.576, while engineering-context and graph-path removal produce the same recovery-classification accuracy. The differences between the accuracy and macro-F1 rankings reflect their different treatment of class frequencies: accuracy summarizes overall correctness, whereas macro-F1 gives equal weight to each predefined class.

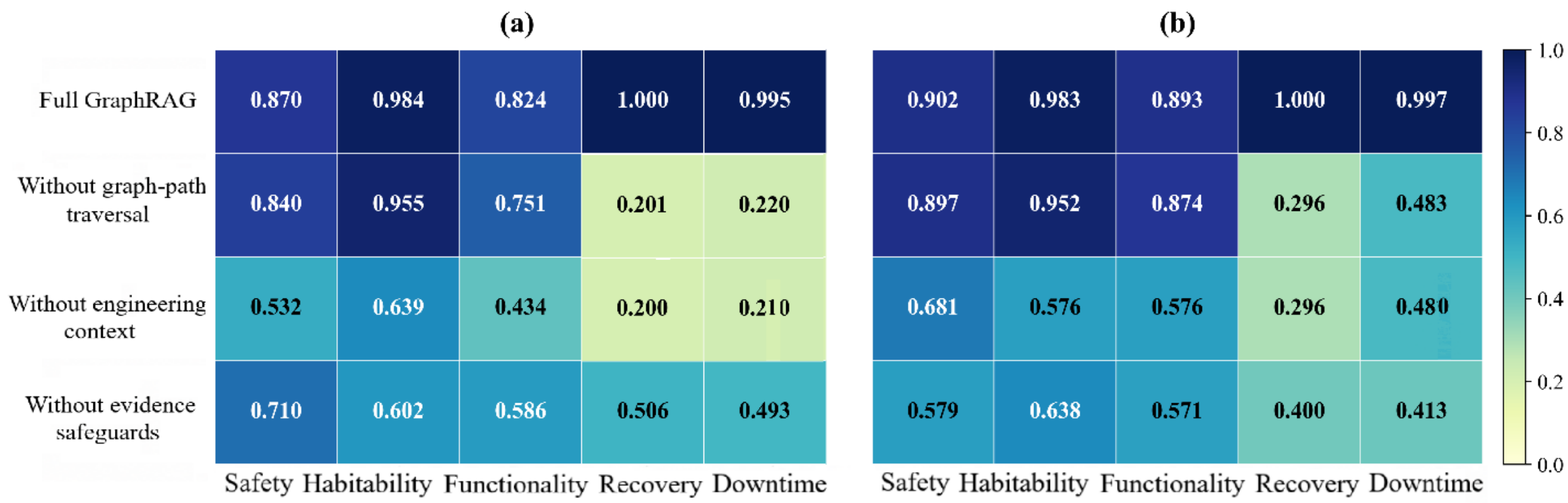

**Fig. 8** Macro-F1 (a) and accuracy (b) of the complete GraphRAG framework and three ablated configurations obtained from the nested leave-one-event-out robustness evaluation.

Across the evaluated configurations, the complete framework provides the most consistent performance, while the relative effect of removing each component varies across outcomes and evaluation metrics. These results indicate that the framework performance arises from the combined use of structured engineering knowledge, controlled evidence handling, and explicit graph relationships rather than from any individual component alone. Because the engineering rules and safeguards were informed by prior corpus diagnostics, the ablation results are interpreted as a robustness and framework-dependence analysis rather than independent validation of the underlying engineering rules.

### 3.6 Sensitivity to input representation

The sensitivity of the framework to the representation of the input evidence was examined using the same damaged residential building described through three alternative pathways: direct image input processed by the implemented vision-language model, an engineering-oriented textual description generated from the image, and an independent engineer-provided description. The comparison evaluates whether differences in the representation of the observed building condition propagate from structured damage extraction to the final building-condition and recovery assessment.

Figure 9 shows the representative building scenario and compares the component-level observations extracted from the three input representations. Differences occur in the described components and assigned damage severities. For example, the engineer description assigns complete roof-structure damage, while Qwen and ChatGPT assign major roof-structure damage. ChatGPT also describes roof-covering damage, whereas this component is not described by the engineer or Qwen. Despite these differences, the three representations lead to the same building-level assessment for safety, habitability, functionality, recovery, and downtime class.

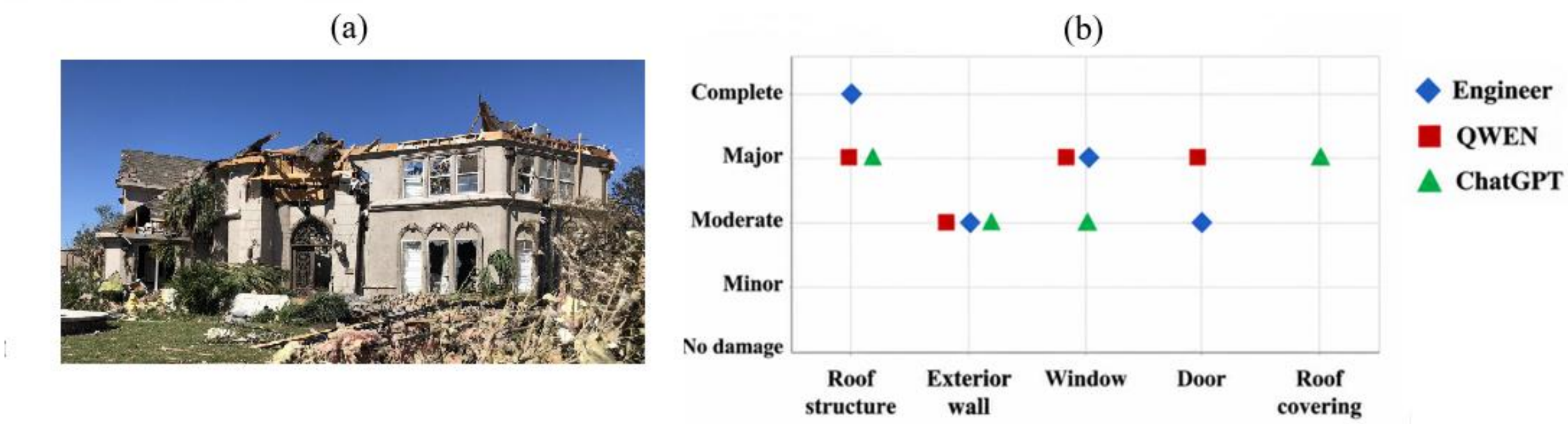

**Fig. 9** Sensitivity to input representation: (a) representative damaged building scenario and (b) component-level damage severity extracted from the three input representations.

The comparison distinguishes variation in evidence extraction from variation in downstream assessment. Where different descriptions lead to similar building-level outcomes, the result indicates that the graph-based reasoning process is relatively robust to moderate differences in evidence representation. Conversely, any differences in final states or recovery estimates identify outcomes that remain sensitive to the completeness or specificity of the input evidence. This analysis also demonstrates the ability of the framework to operate with either direct visual input or sufficiently detailed textual descriptions when images are unavailable.

## 4. Discussion

The proposed framework extends conventional retrieval-augmented generation by embedding post-event building evidence within an explicit engineering knowledge graph that connects observed damage to physical consequences, building-condition states, recovery actions, and downtime. Its main contribution is therefore not simply the retrieval of similar cases, but the combination of multimodal evidence processing, graph-aligned engineering retrieval, evidence-precedence rules, partial-observability safeguards, and recovery-oriented inference within a traceable workflow. The framework also preserves the distinction between directly observed conditions, unresolved evidence, temporary measures, scenario analogues, and fallback engineering knowledge, allowing the resulting assessment to remain linked to the evidence and relationships that support it.

The tornado-residential case study demonstrates how these capabilities operate in practice. The comparison with vector-based RAG shows that the benefit of GraphRAG is not restricted to producing different categorical predictions, but also includes preserving how the assessment may change when unresolved evidence is clarified. Structured BaselineRAG commits to a single analogue-based assessment, whereas GraphRAG retains evidence-dependent branches for the condition concealed by temporary roof protection. Confirmation of roof-covering damage preserves the initial safety, habitability, and functionality states, while confirmation of major localized roof-structure damage results in more restrictive states, major repair, and longer downtime. GraphRAG also retains the engineering paths connecting validated observations to consequences and building states and distinguishes directly supported recovery actions from conditional actions. This capability is particularly important when visible evidence is incomplete or indirect, as in cases involving temporary roof protection or boarded openings.

The aggregate comparison further indicates the contribution of explicit engineering relationships beyond structured evidence representation alone. Because Structured BaselineRAG and GraphRAG use the same standardized case representation, differences between them more directly reflect the additional graph-aligned retrieval and reasoning mechanisms. Across the held-out scenarios, the clearest improvements with GraphRAG are observed for recovery and categorical downtime, for which Macro-F1 increases from 0.918 and 0.721, respectively, to approximately 1.00. Safety and functionality also improve, whereas the difference for habitability is comparatively small, with Structured BaselineRAG and GraphRAG producing approximately the same Macro-F1. Accuracy also increases with GraphRAG for safety, functionality, recovery, and downtime, while habitability remains approximately unchanged. The quantitative comparison should therefore be interpreted together with the case-level demonstrations: predictive agreement is important, but the ability to retain engineering scope, provenance, and uncertainty is an additional property not captured by classification metrics alone.

The ablation analysis provides further evidence that the framework depends on the interaction of several components rather than on graph traversal alone. The relative effect of removing each component varies across the evaluated outcomes and metrics. Removing engineering context produces the strongest deterioration in Macro-F1 for several outcomes, whereas removal of evidence safeguards has a greater effect on accuracy for selected outcomes. Graph-path traversal also contributes to the downstream building-condition and recovery outcomes. This indicates that the graph is most useful when it encodes meaningful engineering structure and is combined with explicit controls governing how evidence can be transferred across cases. In other words, the contribution arises from the combination of structured engineering knowledge, controlled retrieval, and explicit graph relationships, rather than from the use of a graph representation in isolation.

A particularly important feature of the framework is its treatment of partial observability. Post-event photographs rarely provide complete information about every building component, and concealed regions cannot generally be assessed from exterior imagery alone. By distinguishing confirmed damage, explicit negative observations, unresolved conditions, and temporary measures, the framework avoids interpreting missing information as evidence of no damage. Temporary interventions can therefore inform precautionary or recovery actions without automatically establishing the concealed physical condition. This distinction reduces the risk of unsupported assumptions propagating from image interpretation into building-condition or recovery inference.

The input-representation experiment additionally examines whether variations in the description of the same building propagate to the final assessment. For the examined case, the three input pathways produced some differences in component-level damage extraction, particularly in the affected components and assigned severities, but converged to the same building-level outcome: restricted use, not habitable, severely impaired, major-repair recovery, and the same downtime class. This consistency suggests that the framework can tolerate moderate variation in input representation when the engineering information governing the downstream assessment is preserved, while differences that alter decision-relevant evidence could affect the resulting states or recovery estimates. In either case, the experiment is useful because practical post-event information may be supplied through images, reconnaissance text, or engineer-generated descriptions rather than through a single standardized modality.

The recovery component extends the framework beyond damage and functionality assessment by linking supported physical conditions to intervention requirements and downtime. Recovery actions are inferred from accepted damage–consequence–action relationships rather than transferred wholesale from a retrieved scenario, and their duration uncertainty is propagated through an action-dependency network. This provides a transparent connection between observed damage and estimated recovery time. However, these downtime estimates remain engineering-model outputs based on the prescribed recovery-action library, duration distributions, and dependency relationships; they should not be interpreted as empirically validated predictions of actual building-level recovery times.

Several limitations remain. The current implementation is restricted to tornado-affected low-rise residential buildings and primarily exterior post-event evidence, limiting direct generalization to other building types, occupancies, and wind hazards. The quality of the resulting inference also depends on the completeness of the reconnaissance evidence and on the engineering relationships encoded in the knowledge graph. Image-based evidence remains sensitive to viewing angle, occlusion, image quality, and concealed damage. Moreover, the reference safety, habitability, functionality, recovery, and downtime states used in the case study are operational or rule-derived rather than fully independent field observations. The reported performance should therefore be interpreted within the scope of the defined case-study engineering framework.

Future work should evaluate the framework across broader building typologies and wind hazards, incorporate independently observed post-event recovery information where available, and examine how the engineering knowledge base can be expanded without weakening the evidence-control mechanisms established here. The transferability of the framework to other post-disaster assessment settings with rich reconnaissance information, such as earthquake damage, should also be investigated, together with the integration of additional evidence sources such as aerial and satellite imagery. These extensions would help determine whether the traceable damage-to-recovery reasoning demonstrated in the tornado-residential application can generalize to larger and more heterogeneous post-disaster assessment settings.

## 5. Conclusions

This study developed a multimodal engineering Graph Retrieval-Augmented Generation (GraphRAG) framework for post-wind-event building assessment that integrates image- and text-based evidence within a structured, provenance-preserving engineering knowledge graph. The framework links observed component-level damage to physical consequences, building-condition states, recovery actions, and downtime while distinguishing confirmed damage, negative evidence, unresolved conditions, and temporary measures. It supports both graph-grounded retrieval of stored engineering knowledge and inference for previously unseen building cases, with compatibility rules, contradiction handling, evidence precedence, and abstention mechanisms used to limit unsupported inference under partial observability. Demonstrated for tornado-affected low-rise residential buildings, the framework showed the clearest improvements over the non-graph retrieval-augmented baselines for safety, recovery, and categorical downtime, while functionality also improved and habitability performance remained comparable between Structured BaselineRAG and GraphRAG. The case-level comparison further showed that GraphRAG can preserve evidence-dependent assessment branches when the available observations do not fully establish the underlying physical condition. The ablation analysis confirmed that the observed performance depends on the combined contribution of structured engineering context, evidence safeguards, and graph-path traversal rather than on graph representation alone. The proposed framework therefore extends retrieval-augmented generation beyond similarity-based case retrieval by maintaining explicit and traceable connections among observed damage, engineering consequences, assessment states, recovery requirements, and downtime, providing a transparent basis for recovery-oriented post-disaster building assessment.

## Declaration of competing interest

The authors declare that they have no known competing financial interests or personal relationships that could have appeared to influence the work reported in this paper.

## Acknowledgements

This work was supported by the Centre d'études interuniversitaire des structures sous charges extrêmes (CEISCE) and the Natural Sciences and Engineering Research Council of Canada (NSERC) [grant number RGPIN-2022-03492].

## References

[1] Sattar, S., Ryan, K., Arendt, L., Bonowitz, D., Comerio, M., Davis, C., Deierlein, G., and Johnson, K. J. (2021). *Recommended Options for Improving the Built Environment for Post-Earthquake Reoccupancy and Functional Recovery Time*. NIST Special Publication 1254/FEMA P-2090. National Institute of Standards and Technology, Gaithersburg, MD.

[2] Cook, D. T., Liel, A. B., Haselton, C. B., and Koliou, M. (2022). "A framework for operationalizing the assessment of post-earthquake functional recovery of buildings." *Earthquake Spectra*, 38(3), 1972–2007.

[3] Koliou, M., and van de Lindt, J. W. (2020). "Development of building restoration functions for use in community recovery planning to tornadoes." *Natural Hazards Review*, 21(2), 04020004.

[4] Nevill, J. B., and Lombardo, F. T. (2020). "Structural functionality scale for light-framed wood buildings with indicators for windstorm damage." *Journal of Structural Engineering*, 146(4), 04020033.

[5] Nevill, J. B., and Lombardo, F. T. (2021). "Empirical tornado resilience model for light-framed wood residential buildings." *Engineering Structures*, 231, 111840.

[6] Yeum, C. M., Dyke, S. J., Benes, B., et al. (2019). "Postevent reconnaissance image documentation using automated classification." *Journal of Performance of Constructed Facilities*, 33(1), 04018103.

[7] Rodríguez, O., Bech, J., Soriano, J. D., Gutiérrez, D., and Castán, S. (2020). "A methodology to conduct wind damage field surveys for high-impact weather events of convective origin." *Natural Hazards and Earth System Sciences*, 20, 1513–1531.

[8] Roueche, D. B., Nakayama, J. O., Kijewski-Correa, T., & Prevatt, D. O. (2024). A unified multievent windstorm performance testbed for single-family residential buildings. *Natural Hazards Review*, *25*(2), 04023060.

[9] Aghababaei, M., Koliou, M., Pilkington, S., Mahmoud, H., van de Lindt, J. W., Curtis, A., ... & Watson, M. (2020). Validation of time-dependent repair recovery of the building stock following the 2011 Joplin Tornado. *Natural Hazards Review*, *21*(4), 04020038.

[10] Pilkington, S. F., Curtis, A., Mahmoud, H., van de Lindt, J. W., Smith, S., and Ajayakumar, J. (2021). "Preliminary documented recovery patterns and observations from video cataloged data of the 2011 Joplin, Missouri, tornado." *Natural Hazards Review*, 22(1), 05020015.

[11] Molina Hutt, C., Vahanvaty, T., and Kourehpaz, P. (2022). "An analytical framework to assess earthquake-induced downtime and model recovery of buildings." *Earthquake Spectra*, 38(2), 1283–1320.

[12] Wang, W. L., and van de Lindt, J. W. (2021). "Quantitative modeling of residential building disaster recovery and effects of pre- and post-event policies." *International Journal of Disaster Risk Reduction*, 59, 102259.

[13] Wang, W., van De Lindt, J. W., Rosenheim, N., Cutler, H., Hartman, B., Sung Lee, J., & Calderon, D. (2021). Effect of residential building wind retrofits on social and economic community-level resilience metrics. *Journal of Infrastructure Systems*, *27*(4), 04021034.

[14] Gupta, R., Goodman, B., Patel, N., Hosfelt, R., Sajeev, S., Heim, E., Doshi, J., Lucas, K., Choset, H., and Gaston, M. (2019). "Creating xBD: A dataset for assessing building damage from satellite imagery." *Proceedings of the IEEE/CVF Conference on Computer Vision and Pattern Recognition Workshops*, 10–17.

[15] Nex, F., Duarte, D., Tonolo, F. G., and Kerle, N. (2019). "Structural building damage detection with deep learning: Assessment of a state-of-the-art CNN in operational conditions." *Remote Sensing*, 11(23), 2765.

[16] Su, J., Bai, Y., Wang, X., et al. (2020). "Technical solution discussion for key challenges of operational convolutional neural network-based building-damage assessment from satellite imagery: Perspective from benchmark xBD dataset." *Remote Sensing*, 12(22), 3808.

[17] Al Shafian, S., and Hu, D. (2024). "Integrating machine learning and remote sensing in disaster management: A decadal review of post-disaster building damage assessment." *Buildings*, 14(8), 2344.

[18] Braik, A. M., and Koliou, M. (2025). "Post-tornado automated building damage evaluation and recovery prediction by integrating remote sensing, deep learning, and restoration models." *Sustainable Cities and Society*, 123, 106286.

[19] Braik, A. M., Han, X., and Koliou, M. (2025). "A framework for resilience analysis and equitable recovery in tornado-impacted communities using agent-based modeling and computer vision-based damage assessment." *International Journal of Disaster Risk Reduction*, 121, 105427.

[20] Lewis, P., Perez, E., Piktus, A., Petroni, F., Karpukhin, V., Goyal, N., Küttler, H., Lewis, M., Yih, W.-T., Rocktäschel, T., Riedel, S., and Kiela, D. (2020). "Retrieval-augmented generation for knowledge-intensive NLP tasks." *Advances in Neural Information Processing Systems*, 33, 9459–9474.

[21] Karpukhin, V., Oguz, B., Min, S., Lewis, P., Wu, L., Edunov, S., Chen, D., and Yih, W.-T. (2020). "Dense passage retrieval for open-domain question answering." *Proceedings of the 2020 Conference on Empirical Methods in Natural Language Processing*, 6769–6781.

[22] Chen, W., Hu, H., Chen, X., Verga, P., and Cohen, W. (2022). "MuRAG: Multimodal retrieval-augmented generator for open question answering over images and text." *Proceedings of the 2022 Conference on Empirical Methods in Natural Language Processing*, 5558–5570.

[23] Yasunaga, M., Aghajanyan, A., Shi, W., et al. (2023). "Retrieval-augmented multimodal language modeling." *Proceedings of the 40th International Conference on Machine Learning*, 202, 39755–39769.

[24] Miao, J., Lu, D., and Wang, Z. (2025). "A multimodal RAG framework for housing damage assessment: Collaborative optimization of image encoding and policy vector retrieval." *arXiv preprint*, arXiv:2509.09721.

[25] Xiao, Y., Yin, K., and Mostafavi, A. (2026). "CrisiSense-RAG: Crisis sensing multimodal retrieval-augmented generation for rapid disaster impact assessment." *Computer-Aided Civil and Infrastructure Engineering*, 48, 100096.

[26] Yu, Y., Du, X., Hu, Q., et al. (2026). "WeatherArchive-Bench: Benchmarking retrieval-augmented reasoning for historical weather archives." *Proceedings of the 49th International ACM SIGIR Conference on Research and Development in Information Retrieval*, Resource Track. Available as arXiv:2510.05336.

[27] Ghosh, S., and Mittal, G. (2025). "Advancing engineering research through context-aware and knowledge graph–based retrieval-augmented generation." *Frontiers in Artificial Intelligence*, 8, 1697169.

[28] Park, S.-I., and Lee, J.-Y. (2024). "Toward robust RALMs: Revealing the impact of imperfect retrieval on retrieval-augmented language models." *Transactions of the Association for Computational Linguistics*, 12, 1686–1702.

[29] Zhang, Z., Zhu, J., Zhou, W., Qi, X., Zhang, P., and Li, H. (2024). "BoolQuestions: Does dense retrieval understand Boolean logic in language?" *Findings of the Association for Computational Linguistics: EMNLP 2024*, 2767–2779.

[30] Edge, D., Trinh, H., Cheng, N., Bradley, J., Chao, A., Mody, A., Truitt, S., Metropolitansky, D., Ness, R. O., and Larson, J. (2024). "From local to global: A Graph RAG approach to query-focused summarization." *arXiv preprint*, arXiv:2404.16130.

[31] Kuligowski, E. D., Lombardo, F. T., Phan, L. T., Levitan, M. L., and Jorgensen, D. P. (2014). *Technical Investigation of the May 22, 2011, Tornado in Joplin, Missouri*. NIST NCSTAR 3. National Institute of Standards and Technology, Gaithersburg, MD.

[32] StEER Network. (2026). *StEER Network: Building Resilience through Reconnaissance*. https://www.steer.network/. Accessed October 7, 2026.

[33] Roueche, D., Robertson, I., Prevatt, D., Kijewski-Correa, T., and Mosalam, K. (2019). *StEER – 10.20.2019 Dallas, TX EF-3 Tornado: Event Briefing.* DesignSafe-CI.

[34] Sutley, E., Lequesne, R., Li, J., Kirkham, W., Chen, Z., Al-Sabawy, A., Daniel, L., Enderami, S. A., Kim, J., Mudaliar, T., Taher, S., Sharma, P., and Roueche, D. (2019). StEER – 28 May 2019 Linwood, KS EF4 Tornado: Field Assessment Structural Team (FAST) Early Access Reconnaissance Report (EARR). DesignSafe-CI.

[35] Yan, G., Zhang, X., Elgawady, M., Han, D., Li, T., Li, Z., Zhao, Y., Honerkamp, R., Zhao, J., Ramadan, A., Esswein, E. M., Barner, T., and Roueche, D. (2019). *StEER – 22 May 2019 Jefferson City, MO Tornado: Field Assessment Structural Team 1 (FAST-1) Early Access Reconnaissance Report (EARR).*

[36] Wood, R., Roueche, D., Cullum, K., Davis, B., Gutierrez Soto, M., Javadinasab Hormozabad, S., Liao, Y., Lombardo, F., Moravej, M., Pilkington, S., Prevatt, D., Kijewski-Correa, T., Djima, W., and Robertson, I. (2020). *Early Access Reconnaissance Report (EARR), StEER – 3 March 2020 Nashville Tornadoes.* DesignSafe-CI.

[37] Structural Extreme Events Reconnaissance (StEER). (2025). *2025 St. Louis Tornado: Annotated Media Repository*. NHERI DesignSafe, Project ID PRJ-5980.

[38] Kuligowski, E. D., Phan, L. T., Levitan, M. L., and Jorgensen, D. P. (2013). *Preliminary Reconnaissance of the May 20, 2013, Newcastle-Moore Tornado in Oklahoma*. NIST Special Publication 1164. National Institute of Standards and Technology, Gaithersburg, MD.

[39] VanDerostyne, D. A., Hallet, S. K., and Nichols, J. A. (2013). "Post-Event Forensic Investigation of Damaged Structures from Strong Wind Events." *Forensic Engineering 2012: Gateway to a Safer Tomorrow*, ASCE, 841–850.

[40] Kikitsu, H., and Sarkar, P. P. (2015). "Building damage, wind speed estimation, and post-disaster recovery in an EF5 tornado." *Natural Hazards Review*, 16(2), 04014019.

[41] Rhee, D. M., Nevill, J. B., and Lombardo, F. T. (2022). "Comparison of Near-Surface Wind Speed Estimation Techniques Using Different Damage Indicators from a Damage Survey of Naplate, IL EF-3 Tornado." *Natural Hazards Review, 23*(1), 04021052.

[42] Federal Emergency Management Agency. (2019). *Post-Disaster Building Safety Evaluation Guidance*. FEMA P-2055. Federal Emergency Management Agency, Washington, DC.

[43] Applied Technology Council. (2004). *Field Manual: Safety Evaluation of Buildings after Windstorms and Floods*. ATC-45. Applied Technology Council, Redwood City, CA.

[44] Federal Emergency Management Agency. (2021). *Maintenance and Retrofit Options for Roof Coverings, Opening Protection Systems, and Exterior Building Elements*. CNMI MAT Fact Sheet 1 (FS1). FEMA, June 2021.

[45] Federal Emergency Management Agency. (2025). *Individual Assistance Program and Policy Guide*, Version 1.1, Amended July 2025. FEMA.